\documentclass[prl,twocolumn,superscriptaddress,showpacs,longbibliography]{revtex4-1}
\usepackage{graphicx,amsmath,amssymb,bm}
\usepackage{comment}
\usepackage{multirow}
\usepackage{xcolor}
\usepackage{hhline}
\newcommand{\be}{\begin{equation}}
\newcommand{\ee}{\end{equation}}

\newcommand{\fm}{\, \text{fm}}
\newcommand{\mev}{\, \text{MeV}}
\newcommand{\kev}{\, \text{keV}}

\newcommand{\NNLO}{\text{N}^2\text{LO}}

\newcommand{\pilesseft}{\mbox{$\pi\text{\hspace{-5.5pt}/}$}EFT$\,$}

\newcommand{\cblack}{\color{black} }

\graphicspath{{./images/}}
\begin{document}

\title{Theoretical evaluation of solar proton-proton fusion reaction rate and its uncertainties}

\author{Hilla\ De-Leon}
\email[E-mail:~]{hilla.deleon@mail.huji.ac.il}
\affiliation{Racah Institute of Physics, 
The Hebrew University, 
9190401 Jerusalem, Israel}
\affiliation{INFN-TIFPA Trento Institute of Fundamental Physics and Applications, Via Sommarive, 14, 38123 Povo TN, Italy}
 \affiliation{
 European Centre for Theoretical Studies in Nuclear Physics and Related Areas (ECT*),
 Strada delle Tabarelle 286, I-38123 Villazzano (TN), Italy}
\author{Doron\ Gazit}
\email[E-mail:~]{doron.gazit@mail.huji.ac.il}
\affiliation{Racah Institute of Physics, 
The Hebrew University, 
9190401 Jerusalem, Israel}

\begin{abstract}
The weak proton-proton fusion into a deuteron ($^2$H) is the driving reaction in the energy production in the Sun, 
as well as similar main sequence stars. Its reaction rate in the solar interior is determined only theoretically. 
Here, we provide a new determination of the rate of this reaction in solar conditions $S^{11}(0)$, 
and analyze theoretical and experimental sources for uncertainties, using effective field theory of 
quantum chromo-dynamics without explicit pions at next-to-leading order. 
We find an enhancement of $S^{11}$ by $1-4\%$ over the previously recommended value. 
This change reduces the calculated fluxes of neutrinos originating in $^8$B and $^7$Be nuclear reactions in the Sun, 
thus favoring higher abundances for metallic photospheric elements, in the tension between different composition determination, known as the ``Solar Composition Problem''.
\end{abstract}

\pacs{}
\keywords{}
\maketitle

The evolution of the Sun remains one of the main theoretical questions in astrophysics. 
The properties of the Sun, due to its proximity to earth, are measured to high precision using a multitude of elementary messengers. 
Combined with the theoretical modeling of solar evolution, the Sun is considered and used as a laboratory. This approach has led in the past 
to the discovery of a finite neutrino mass, and, consequently, neutrino flavour oscillations, inferred from a deficiency between measured and expected neutrino fluxes \cite{PhysRevLett.78.171, Bahcall_2001}.

The energy generated in the Sun originates in an exothermic set of reactions, 
the proton-proton ($pp$) chain, by fusing four protons into $^4$He nucleus. 
In all but a few per-mils of the reactions, the chain is initiated by a 
weak proton-proton ($pp$)-fusion into a deuteron, $p+p\rightarrow d+e^++\nu_e$ \cite{Borexino_2018}. 
The solar conditions, whose extreme temperature is low compared to the needed kinetic energy for the protons to overcome 
the Coulomb repulsion, make this reaction highly improbable, with a half-life of about $1$ billion years, 
determining the time scale of the energy production of the Sun. The importance of this reaction is also reflected in the high
sensitivity of other solar observables, such as neutrino spectra from $^8$B and $^7$Be reactions, to the $pp$-fusion rate~\cite{Vinyoles:2016djt}.

This slow rate and low characteristic energy also make a laboratory measurement of the cross-section challenging, 
thus currently, it is calculated theoretically. The aforementioned importance of this reaction for solar modeling entails a need for precise and accurate
calculation, with an uncertainty objective of $\approx 1\%$. 
As we detail hereafter, many theoretical studies were accomplished in the last decade~\cite{solar1, PhysRevLett.110.192503, Park:2002yp, Acharya:2016kfl, Proton_Proton_Fifth_Order, Ando_proton, PhysRevLett.119.062002,chen_proton}, that differ in their uncertainty estimate, and with a general trend of increasing the previously accepted value~\cite{RevModPhys.83.195}, thus creating a challenge in understanding the consequences on solar observables. 
\cblack
In the current {\it letter}, we present a novel calculation of the proton-proton fusion rate,
uniquely developing an approach whose most important property is the possibility to robustly and reliably assess the systematic theoretical uncertainty. 

The $pp$ fusion cross-section, $\sigma_{pp}=\frac{S^{11}(E)}{E}\exp[-2\pi\eta(E)]$, is 
separated to a long-range Coulomb suppression factor ($\eta(E)$), and a 
short-range nuclear dependent part, {\it astrophysical $S$-factor} ($S^{11}(E)$), 
where $E$ is the kinetic energy of the center of mass of the interacting protons, 
dictated by the temperature of the core~$1.5\kev$. At these low energies, compared to nuclear characteristic energies, $S^{11}(E)$
is dominated by the $E=0$ threshold value, \textit{i.e.}, $S^{11}(0)$, which is conveniently parameterized in the following way~\cite{RevModPhys.83.195}:
\begin{equation}\label{eq_S_11}
\begin{split}
&S^{11}(0)=(4.011\pm0.04)\cdot 10^{-23}\mev\cdot\fm^2 \cdot\\ &\left(\frac{(ft)_{0^+\rightarrow0^+}}{3071.4 \text{sec}}\right)^{-1}\left(\frac{g_A}{1.2695}\right)^2\left(\frac{f^R_{pp}}{0.144}\right)\left(\frac{\Lambda_{pp}^2}{7.035}\right).
\end{split}
\end{equation}
Here, $\Lambda_{pp}^2$ is the square of the $pp$-fusion matrix element, whereas $g_A$ is the axial charge of the nucleon, $(ft)_{0^+\rightarrow0^+}$ is the value for
superallowed $0^+\rightarrow0^+$ transitions that has been determined from a comprehensive analysis of experimental
 rates corrected for radiative and Coulomb effects \cite{PhysRevC.79.055502}. $f^R_{pp}$ is a phase-space factor, that also consider 1.62\% increase due to radiative corrections to the cross-section \cite{PhysRevC.67.035502}.
The current recommended value of the axial charge of the nucleon, $g_A=1.2756\pm 0.0013$~\cite{gA_2020}, represents a change in value and increased uncertainty compared to past extractions, stemming from a 
current tension between different neutron half-life measurements \cite{Markisch:2018ndu,doi:10.1063/1.4983578}. 

The main challenge in theoretically determining the {\it{S-factor}} is evaluating the nuclear matrix element $\Lambda_{pp}^2(0)$,
as it originates in the non-perturbative character of the fundamental theory, Quantum Chromodynamics (QCD), at the nuclear regime. This makes a direct calculation non-trivial, nonetheless estimating the 
theoretical uncertainty since the calculation usually includes an uncontrolled approximation due to a choice of the model used to describe the nuclear forces and currents.

The recent comprehensive review of cross-sections of nuclear reactions relevant to solar evolution, SFII \cite{RevModPhys.83.195}, which is the origin of Eq.~\ref{eq_S_11}, has recommended using the value $\Lambda_{pp}^2(0)=7.035$, with an uncertainty of about $1\%$, 
that was taken from the difference in the results of three different theoretical approaches. 
This error estimation is clearly an uncontrolled approach to error estimates. The modern approach to nuclear physics treats the nuclear interaction as a low energy effective field theory (EFT) of QCD and lays out a model-independent path to estimate systematic theoretical uncertainties. 
EFT provides a renormalizable and, in principle, a model-independent theoretical method for describing low energy reactions in cases where the underlying fundamental theory is non-perturbative. EFT takes advantage of energy scale separation between typical momentum in the process, $Q$, is small compared to a physical cutoff, $\Lambda_{cut}$ ({\it, i.e.,} $Q/\Lambda_{cut}\ll 1$), to systematically expand forces and scattering operators. This scale separation enables theoretical uncertainty assessment through the convergence rate of the expansion. 

The typical momentum of a nuclear reaction is determined by the momentum transfer and the typical momentum of the nuclei involved. The momentum transfer during the $pp$-fusion, $Q \sim \text{few keV}$, is much smaller than even the shallowest QCD excitation, namely the pion. Moreover, the momentum scales in light nuclear reactions, i.e., the nucleon-nucleon singlet scattering length, $a_s -23.714 \fm$, and the typical binding momenta of the $A=2,\,3$ nuclei ($\gamma_{typical}=\sqrt{2M_N E/A}$, where $M_N$ is the nucleon mass and $E$ is the binding energy), 
 $\gamma_d \approx 45\mev$, and $\gamma_{^3\text{H},\, {^3\text{He}}}\approx 70-75\mev$, 
are all significantly lower than the pion excitation. This means that, in principle, pion-less EFT (\pilesseft), where all degrees of freedom, but the nucleons, are integrated out, and their properties dictate the size of the effective Lagrangian coefficients, called low-energy constants (LECs), can be used to calculate $pp$-fusion \cite{few_platter, Griesshammer_pionless, few, Kaplan1996629, KSW1998_a, Kaplan:1998tg, KSW_c, Hammer:2019poc}. Such theories can be easily made renormalizable, but their small applicable energy regime creates a problem in calibrating the EFT parameters. Particularly, the strength of interaction between coupled nucleons and a weak probe, which at next-to-leading-order (NLO) involves a new low-energy constant, $L_{1, A}$, that needs to be calibrated \cite{Proton_Proton_Fifth_Order, Ando_proton, Kong2}. A recent calculation by NPLQCD collaboration used Lattice QCD to directly assess $L_{1, A}$ \cite{PhysRevLett.119.062002}, albeit limited to non-physical pion mass. They then employed it in a \pilesseft calculation to estimate $pp$-fusion rate while extrapolating to physical pion mass. Thus, still plagued with possible unaccounted for systematic uncertainties, especially considering the aforementioned precision needs \cite{cirigliano2022neutrinoless,beane2009high,parisi1984strategy,lepage1989analysis}. 

To avoid these challenges of \pilesseft, state-of-the-art calculations~\cite{PhysRevLett.110.192503,Acharya:2019zil,Acharya:2016kfl} use chiral EFT,
i.e., a non-relativistic EFT of nucleons and pions. Chiral EFT breakdown energy is above the pion excitation; thus, it is applicable 
to describe nuclear properties for all bound nuclei or low-energy reactions. In particular, weak reactions involving nuclei of masses as high as $A\approx 100$ have been calculated {\textit{ab-initio}} \cite{Gysbers:2019uyb}, with accuracy 
reaching a few percent, even for these nuclei. 
However, Chiral EFT calculations include many parameters, turning the calibration of its forces and scattering operators into 
a complicated statistical problem \cite{PhysRevX.6.011019, Acharya:2016kfl}, which is also the source of the uncertainty estimate. Additionally, since current formulations are nonrenormalizable, regularization and nonrelativistic reduction create inconsistencies and inherent model dependence, with unknown consequences on the accuracy of predictions. Moreover, chiral EFT calculations still mix between different EFT orders in the two- and three-body forces and the scattering operator. Finally, these calculations have used a mistaken relation between the weak coupling strength and the three-body force, \cite{PhysRevLett.103.102502, PhysRevLett.122.029901}, which is expected to change the nominal result upwards by about $0.5\%$~\footnote{B.~Acharya, private communication}. Within these caveats, and in the context of $pp$-fusion, both direct calculations \cite{Acharya:2019zil,Acharya:2016kfl} and chiral EFT estimates of $L_{1,A}$ \cite{Acharya:2019fij} appeared in the recent literature. 

Fig.~1 compares the $pp$-fusion rate inferred from these calculations to the recommended value~\cite{RevModPhys.83.195}. A general trend of increase from the recommended value appears.

Motivated by these, we present here a novel approach, based on \pilesseft calculations, to predict the $pp$-fusion rate. We first use the $^3$H $\beta$ decay rate to calibrate $L_{1, A}$ for various energy scales. We then use it to calculate the $pp$-fusion nuclear matrix element. Concomitantly, we use a similar approach for analogue electromagnetic reactions~\cite{de2020shell} to verify and validate the procedure. Specifically, we calculate the $^3$H and $^3$He magnetic moments. These are used to calibrate an NLO low-energy-constant named $L_1$, coupling a pair of nucleons to a photon and governing 
the strength of a transition between spin-singlet and spin-triplet coupled pair of nucleons, similar to the weak $L_{1, A}$. $L_1$ is then employed to ``post-dict'' the $np$-fusion nuclear matrix element. The entire procedure is conveniently depicted in Fig.~\ref{fig_explanation}.
This prediction method includes several key ingredients chosen to ensure a robust uncertainty estimate. 
First, we show that the theory is renormalizable. Second, we use a systematic order-by-order analysis, show that the perturbative, next-to-leading order (NLO) is indeed much smaller than the leading order (LO), and use this fact to estimate theoretical uncertainty. 
Third, the calculation consists of a small number of parameters, all of which have a physical interpretation, thus minimizing statistical uncertainty. Last, we validate and verify the theory, and its uncertainty estimate, using well-measured analogue transitions.
 
 \begin{figure}[h!]
 \label{fig_S11}
 \includegraphics[width=1\linewidth]{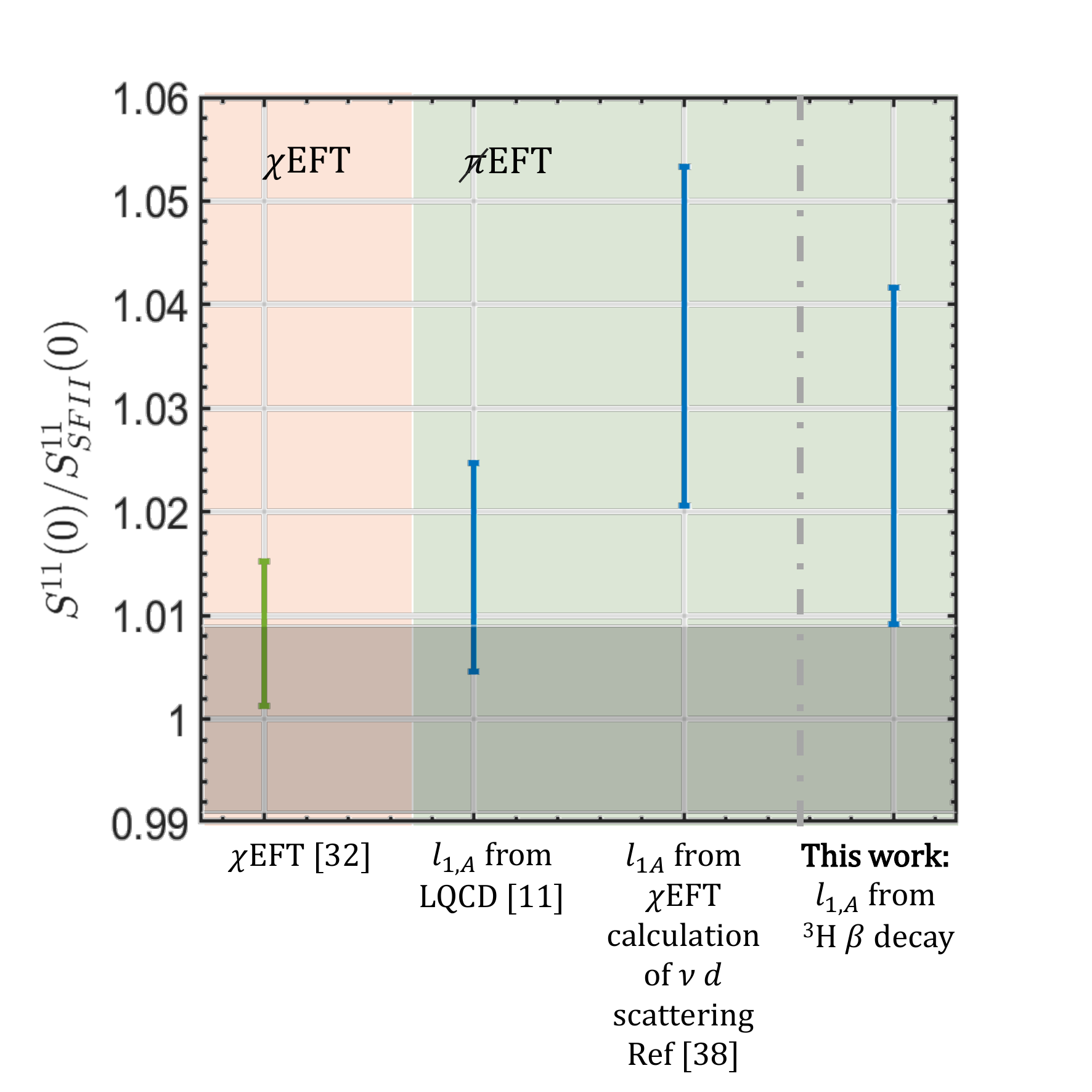}
 \caption{\footnotesize{\textbf{Recent $S^{11}(0)$ calculations compared to SFII recommended value~\cite{RevModPhys.83.195}}. The gray band is the SFII uncertainty estimate. $\chi$EFT calculation is denoted by the light red background~\cite{Acharya:2019zil}.
 \pilesseft calculations with different $l_{1,A}$ calibrations have a green background, Refs.~\cite{PhysRevLett.119.062002,Acharya:2019fij} and the current work. We remark that the value plotted as Ref.\cite{Acharya:2019fij} infers $S^{11}(0)$ from the $l_{1,A}$ calculated in that reference, combined with Eqs.~\ref{eq_S_11} and \ref{eq:Lambda_pp_calc} from that letter.}}
 \end{figure}

The calculation uses \pilesseft expansion to NLO, consistently in the forces and the scattering operators. The operator similarity between Gamow-Teller transitions and iso-vector magnetic moments has been used in traditional nuclear physics. This is due to the fact that both these reactions include a spin and isospin coupling, which at the single particle approximation correspond to spin ($\vec{\sigma}$) and isospin ($\vec{\tau}$) Pauli matrices combination $\vec{\sigma}\tau^{(a)}$, with the isospin component being $a=0$ for iso-vector magnetic moment, and $a=\pm 1$ for Gamow-Teller $\beta$-decays. This analogy breaks when considering the interaction of the different probes with higher energy degrees of freedom, like the pion or delta.
The fact that the strong interaction is almost isospin invariant and that the Coulomb repulsion between the protons is a minor effect, strengthens this analogy, as it entails a proximity between the wave functions of isobaric analogues. This, in turn, minimizes the difference between magnetic moments of these analogues and transitions between them, as induced by the Gamow-Teller operator.

Moreover, in \pilesseft, this analogy continues from the single nucleon level (at LO) to higher orders. In Tab.~\ref{tab:EM_Weak_comparison} we directly compare the structure of the nuclear non-conserved currents, i.e., magnetic and weak-axial, up to NLO. The currents are evidently identical in their structure, apart from the isoscalar parts in the magnetic current, as well as energy scale dependence (renormalization group behavior). Since scattering operators are derived from these currents, the operators for iso-vector are identical to the weak axial operators. This is the reason why in \pilesseft, quantitatively study of the magnetic structure of isobaric analogues, and the inherent theoretical uncertainty due to neglected higher orders contributions, can be used to verify and validate the Gamow-Teller transitions. The calculations we present in the following are made feasible by applying recent advancements in \pilesseft, allowing the evaluation of matrix elements between three-nucleon states, fully including the Coulomb interaction between protons, needed to calculate the properties of $^3$He~\cite{De-Leon:2019dqq, Konig:2015aka,konig2,konig1, konig5, konig3}. In a preceding paper, we have used this to calculate $^3$H $\beta$-decay \cite{de2019tritium} and in an article accompanying the current {\it{letter}}, we use the theory to calculate zero momentum transfer electromagnetic observables of 2 and 3 nucleonic states~\cite{de2020shell}. For both calculations, to eliminate any artificial effects, we choose to use the same value $\left(\mu=
 \Lambda\rightarrow \infty\right)$ for both the two-body dimensional regularization and the three-body cutoff regularization (see also Ref.~\cite{Hammer:2019poc}).

Furthermore, for $A=3$ mass nuclei, the Coulomb interaction between protons in $^3$He minimally changes the difference from $^3$H, the isobaric analogue. This can be seen by studying the matrix element of the Fermi operator, i.e., the isospin changing operator (interchanging a neutron in $^3$H into a proton in $^3$He) between these states. In Fig.~\ref{fig:EW_observables}, the breakdown scale dependence, i.e., the Renormalization Group behavior, of this matrix element is shown. Markedly, the wave-functions of the $A=3$ isobaric analogues are identical, up to $0.1-0.2\%$ deviation due to the Coulomb interaction, compared to the zero momentum transfer electric charge of $^3$H and $^3$He, i.e., $\langle^3\text{H}|{\tau}^0|^3\text{H}\rangle=\langle^3\text{He}|{\tau}^0|^3\text{He}\rangle=1$ (see Refs.\cite{de2019tritium,De-Leon:2019dqq}).

 \begin{table*}[ht!]
 \centering
 \begin{tabular}{c|c|c}
 &Nuclear non-conserved current&two-body contact strength \\
 \hhline{=|=|=}
 Electromagnetic &$A_{i}=\frac{e}{2M}\Big\{\underbrace{N^\dagger\left[\left(\kappa_0+\kappa_1\tau_3\right){\sigma_i}\right]\ N}_{\text{LO}}-\underbrace{2\kappa_0 (\vec{t}^\dagger \times \vec{t})_i}_{\text{NLO}} $&\multirow{2}{*}{$L_1'(\mu)=-\frac{\rho_t+\rho_C}{\sqrt{\rho_t\rho_C}}\kappa_1$+$\underbrace{\frac{M}{\pi\sqrt{\rho_t\rho_C}}L_{1}\left(\mu-\gamma_d\right)\left(\mu-\frac{1}{a_{s}}\right)}_{l_{1},\text{ RG invariant combinations}}$}\\
 nuclear observables&$-\underbrace{L_1'(t_i^\dagger s+s^\dagger t_i)}_{\text{NLO}}\Big\}$&\\
 \hline
 Weak nuclear &\multirow{2}{*}{$A_i^{\pm}=\underbrace{\frac{g_A}{2}N^\dagger\sigma_i\tau^{\pm}N}_{\text{LO}} -\underbrace{L'_{1, A} \left (t_i^\dagger s+s^\dagger t_i\right)}_{\text{NLO}}$}&
 \multirow{2}{*}{$L_{1,A}'(\mu)=-\frac{\rho_t+\rho_C}{2\sqrt{\rho_C\rho_t}}g_A$+$\underbrace{\frac{1}{2\pi\sqrt{\rho_C\rho_t }}L_{1, A}\left(\mu-\gamma_d\right)\left(\mu-\frac{1}{a_s}\right)}_{l_{1,A},\text{ RG invariant combinations}}$}
 \\
 observables &&
 \end{tabular}
 \vspace{0.3 cm}
 \caption{\footnotesize{\textbf{The similarity between the weak and the electromagnetic currents and low energy constants (LEC's)}.
 $A_i(A_i^{\pm})$ is the electromagnetic (weak) current up to NLO, which contains one- and two-body parts.
 For the electromagnetic current, the one-body LECs are $\kappa_0 (\kappa_1) $ the nucleon isoscalar (isovector) magnetic moment. For the weak current, the one-body LEC is $g_A$. 
 For the two-body current, $l_1\,(l_{1,A})$ is an RG invariant combination of the two-body low-energy constant, $L_1\,(L_{1,A}).$ $t(s)$ is an auxiliary field of two coupled nucleons to triplet (singlet) spin state \cite{rearrange}, $\rho_t(\rho_C)$ is the triplet (singlet) effective range and $\mu$ is the renormalization scale.}}
 \label{tab:EM_Weak_comparison}
 \end{table*}

\begin{figure}[t]
 \centering
 \includegraphics[width=1\linewidth]{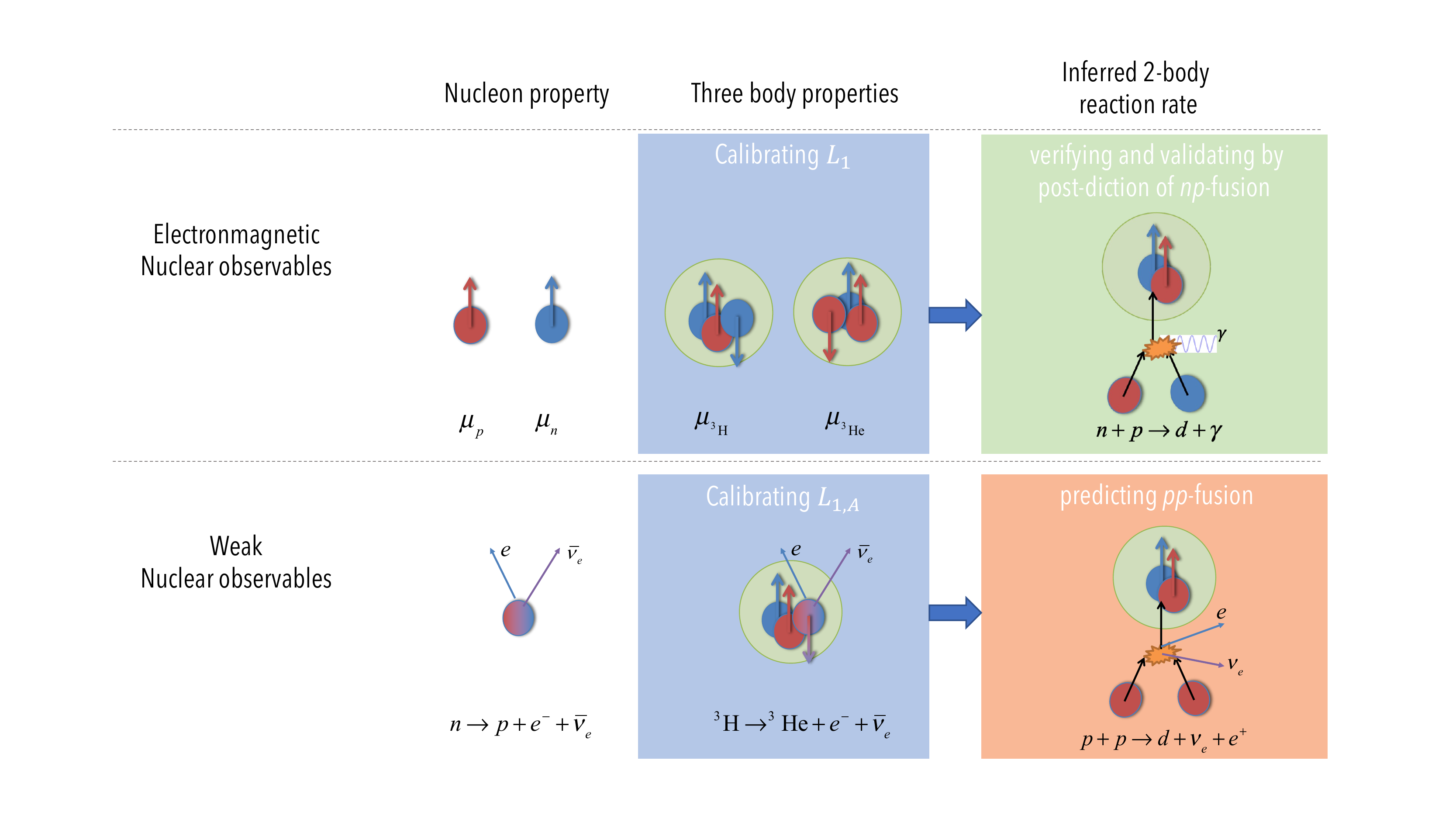}
 \caption{ \label{fig_explanation}\footnotesize{\textbf{The approach used in this work to predict the $pp$-fusion rate.} 
 Top panel: Electromagnetic
nuclear observables; Bottom panel: Weak
nuclear observables. For both panel: The single nucleon and three-body observables (Electromagnetic: $\mu_{^3\text{H}}$ and $\mu_{^3\text{He}}$ which are used to calculate the $np\rightarrow d +\gamma $ rate \cite{de2020shell}; Weak: $^3\text{H}$ $\beta$ decay) are used to calculate the continuum two-body fusion. A successful comparison of the calculated $np\rightarrow d +\gamma $ to experiment then validates the weak $p+p \rightarrow d+e^+ 
 \nu_e$, the $pp$ fusion, prediction.
}}

 \end{figure}

The transition matrix element of $pp$-fusion (see \cite{Proton_Proton_Fifth_Order, Ando_proton}), when terms are limited consistently to NLO (without any higher order terms), is:
\begin{equation} \label{eq:Lambda_pp_calc}
 \begin{split}
 |\Lambda_{pp}(0)|=&\sqrt{Z_d^{NLO}}\left[e^\chi -M\alpha a_CI(\chi)\right]+\\
 &+\sqrt{Z_d^{LO}}{a_C\gamma_d^2}\left(\dfrac{\rho_t+\rho_C}{4}-\frac{1}{2}\sqrt{\rho_t\rho_C}l_{1,A}\right)
 \approx\\
 &\approx 2.655+0.6\times l_{1,A}\approx 2.685
 \end{split}
 \end{equation}
 where $Z_d$ is the deuteron residue, and
 $I(\chi)=\int dx\frac{\chi e^{\frac{x}{\pi} \arctan\left(\frac{\pi \chi}{x}\right)}}{\left(e^x-1\right) \left(\pi^2\chi^2+x^2\right)}$~,
 $\chi=\frac{\alpha M}{\gamma_d}$, and $\alpha\approx\frac{1}{137}$ is the fine-structure
 constant
 \cite{Kong1,Kong2}, $\rho_t(\rho_C)$ is the triplet (singlet) effective range and $a_C$ is the $pp$ scattering length. The final equality is achieved upon taking the central value $l_{1, A}=0.051$
 from our NLO calculation of the $^3$H $\beta$-decay~\cite{de2019tritium}.
The uncertainty of this calculation is discussed in detail hereafter.
Similarly, the cross-section of thermal neutrons ($q=0.0069 \mev/c$) capture by protons $n+p\rightarrow d+\gamma$ is related to the nuclear matrix element $\sigma_{np}\propto |Y'_{np}|^2$. In fact, the only difference in the diagrammatic representation of the $pp$- and $np$-fusion processes is that the latter does not include Coulomb interaction between the interacting nucleons at the continuum's initial state. The resulting formula up to NLO is indeed very similar to Eq.~\ref{eq:Lambda_pp_calc}~\cite{Vanasse:2017kgh,PhysRevLett.115.132001}:
\begin{eqnarray} \label{eq_Y_fineal}
\nonumber Y'_{np} &=&\left (1-\frac{1}{\gamma_da_s}\right) \cdot \sqrt{Z_d^{\text{NLO}}}-\frac{\gamma_d}{4}\left[\rho_t+\rho_C+ \frac{\sqrt{\rho_t\rho_C}}{\kappa_1}l_1 (\mu)\right]
 \\ &\approx& 1.206+0.07 \times l_{1}~,
 \end{eqnarray} 
where $l_{1}$ is an RG invariant combination proportional to $L_1$, the two body iso-vector magnetic LEC~\cite{de2020shell}.

\begin{figure}[h]
 \includegraphics[width=1\linewidth]{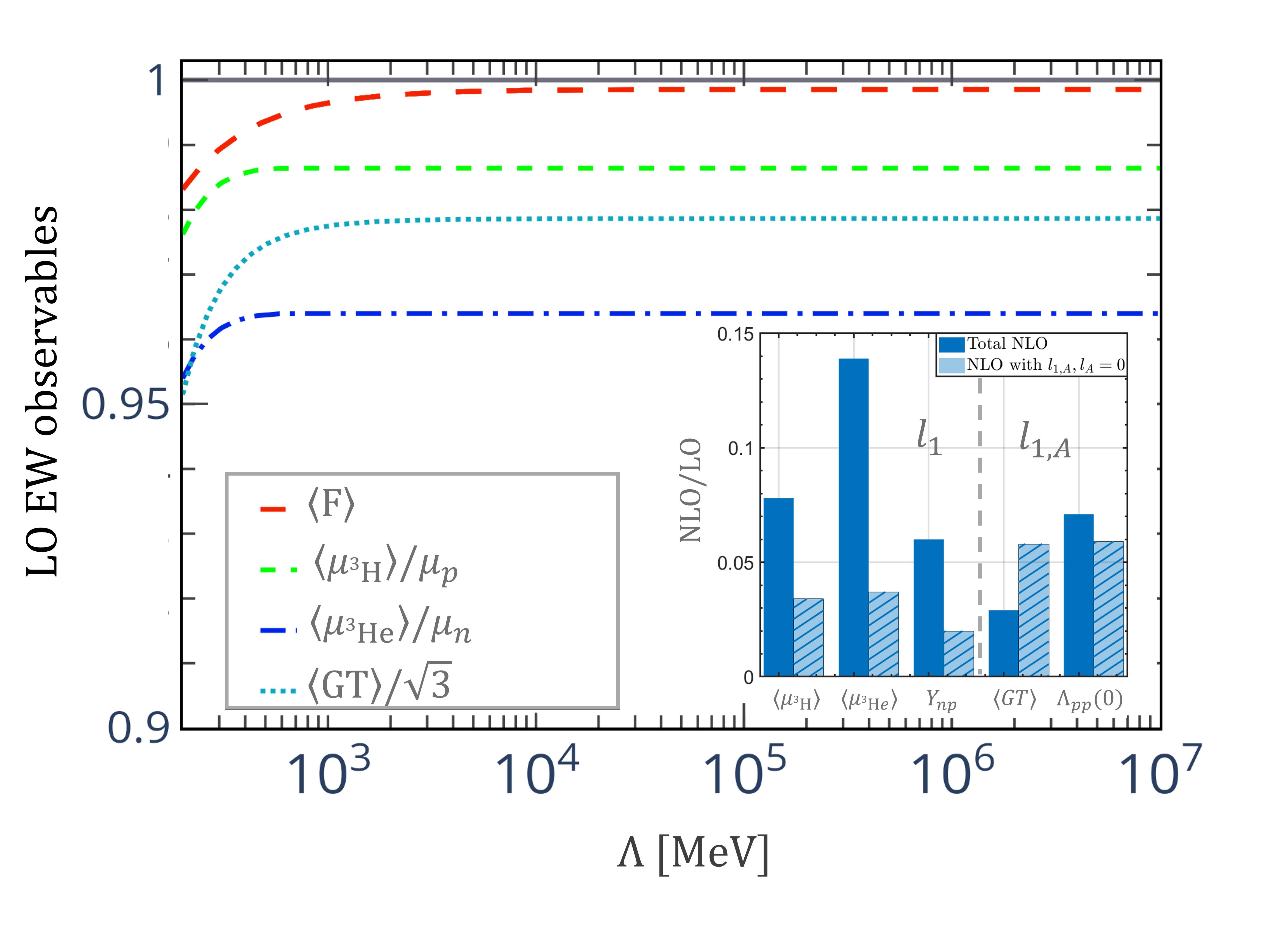}
 \caption{\label{fig:EW_observables}\footnotesize{\textbf{The LO and NLO electroweak (EW) observables calculated in Refs.~\cite{De-Leon:2019dqq,de2019tritium,de2020shell} and this work}. Outer graph: The energy scale dependence (Renormalization group behavior) of LO observables calculated in Refs.~\cite{De-Leon:2019dqq,de2019tritium,de2020shell}, long dashed line: $^3$H-$^3$He Fermi transition; short (long) dashed-dotted line: $^3$H ($^3$He) magnetic moments. Dotted line: GT transition. 
 Inset: The NLO contributions to the weak and electromagnetic observables discussed in the text \cite{De-Leon:2019dqq,de2019tritium,de2020shell}. 
 solid band: the NLO to LO contribution.; striped band: the long-range NLO to LO ratio, i.e., assuming $l_1,l_{1,A}=0$.}}
 \end{figure}
In the electromagnetic sector, the resulting theoretical matrix element and its uncertainty, as taken from Ref.~\cite{de2020shell}, are $Y'_{np}=1.253\pm 0.006$ (the origin of the uncertainty estimate is explained below). This is to be compared with the experimental value of $Y'^{exp}_{np} = 1.2532 \pm 0.0019$~\cite{np_data}. This shows the predictive strength of the theory, to a precision and accuracy of $0.5\%$ for this matrix element. 

The validity of the theory is strengthened when studying, in Fig.~\ref{fig:EW_observables}, the energy scale dependence of the $A=3$ nuclear matrix elements and showing Renormalization group invariance, as manifested in the stabilization of the calculated matrix elements at energy scales of the order of a few times the pion mass~\footnote{We note that the Fermi function stabilizes at higher energy than the other matrix elements. This is due to the Coulomb repulsion between protons, whose physics is of a shorter range than the pion range}. In \pilesseft, the RG behavior of two- and three-body systems is very different. The latter requires the addition, already at leading order (LO), of a 3-body force counter-term, whose strength is dependent on the cutoff $\Lambda$, $H(\Lambda)$~\cite{3bosons, Triton}. Moreover, the NLO contributions to the different observables, shown in the inset of Fig.~\ref{fig:EW_observables}, are small, demonstrating the perturbative nature of the calculation and establishing the order-by-order ratio as a basis for quantifying uncertainty. It is interesting to see that the contribution of $l_{1}$ is dominant in the NLO contribution, whereas that of $l_{1, A}$ is quite smaller than the long-range NLO contribution. We suspect that this is reminiscent of the fact that in chiral EFT, which captures the more fundamental pion physics, the electromagnetic two-body current, i.e., the origin of $l_1$, arises at NLO, while the weak two-body current is suppressed to higher orders \cite{PhysRevLett.114.082502}.

Estimating theoretical uncertainty as the truncation error in an EFT expansion has been a main research topic in recent years, e.g., Refs.~\cite{Griesshammer:2015ahu, PhysRevC.92.024005,Cacciari:2011ze}. Here, we use the procedure used in Ref.~\cite{de2020shell}, to establish the theoretical uncertainty of the magnetic structure of $A=2,\,3$ mass nuclear systems, using \pilesseft at NLO, which is based on the aforementioned references.

Up to NLO, the \pilesseft expansion for any $M_1$ observable observables can be written as:
\begin{equation}
\langle M_1\rangle=\langle M_1\rangle_\text{LO} \left (1+\alpha_1 \cdot \delta_{M_1}+\mathcal{O}(\delta_{M_1}^2)\right).
\end{equation}
Here, $\alpha_1 \cdot \delta_{M_1}$ is the ratio of NLO to LO contributions, $\delta_{M_1}$ is the expansion parameter of the EFT expansion, and $\alpha_1$ is the expansion coefficient that, if not suppressed due to symmetry considerations, should be of order $1$. The truncation error, as in any power-series expansion, should be $\mathcal{O}(\delta_{M_1}^2)$. Refs.~\cite{Griesshammer:2015ahu, PhysRevC.92.024005} use a Bayesian approach to quantify its size, using $\alpha_1$, given the size of $\delta_{M_1}$. However, the expansion parameter is also known up to a factor of the order of $1$. Thus, in Ref.~\cite{de2020shell}, we use NLO to LO ratios of different observables calculated within the same EFT and, in a Bayesian way, estimate the size of the expansion parameter and the truncation error simultaneously. 

As discussed aforementioned, the different magnetic and weak observables are all described using the same EFT; thus, we expect them to share expansion properties. This expansion parameter was estimated in Ref.~\cite{de2020shell}, studying \pilesseft magnetic structure observables. That study concluded that $\delta_{magnetic} \approx 0.05-0.1$. We expect the NLO corrections $^3$H Gamow-Teller strength and $pp$ fusion to be of the same relative size. In the inset of Fig.~\ref{fig:EW_observables} the relative corrections are plotted. Indeed, the NLO contributions of the weak contributions follow the expected expansion trend, and a Grubbs test showed that none of the NLO corrections could be regarded as an outlier.

Further using the uncertainty estimate procedure introduced in Ref.~\cite{de2020shell}, we claim that using $l_{1,A}$ extracted from $^3$H $\beta$ decay \cite{de2019tritium}, which results in a 7\% NLO correction to the $pp$-fusion rate, leads to a theoretical uncertainty of 0.8\% (with 70\% degree of belief) in the nuclear matrix element: $ \Lambda_{pp}^2(0)=(7.21\pm 0.03_{g_A} \pm 0.005_{exp}\pm 0.11)$, with uncertainties due to $g_A$ (see supplementary materials), experimental uncertainties in the triton half-life , and the theoretical uncertainty calculated above, respectively.

Summarizing, in this {\it letter} we have presented a \pilesseft prediction of the $pp$-fusion rate up to next-to-leading-order using weak low energy constant $L_{1, A}$ that is calibrated, for the first time, using a consistent calculation of the $^3\text{H}\rightarrow ^3\text{He}+e^- + \bar{\nu}_e$, $\beta$ decay. The calculation is found to be renormalization group invariant, with small NLO corrections. The calculation is verified and validated by noticing an analogy between the weak sector and the magnetic structure of these systems. Through the use of the magnetic moments of the three body nuclei, a low-energy constant $L_1$ is determined, which indicates the strength of the iso-vector interaction between two coupled nucleons. Hence, a precise and accurate prediction of the radiative thermal neutron capture on a proton is accomplished. Since this is the electromagnetic reciprocal reaction to $pp$ weak fusion, this post-diction and analogy verify the $pp$ fusion calculation. Augmented with a Bayesian approach to robustly estimate the theoretical uncertainty due to truncation error originating in the EFT expansion, this procedure removes concerns regarding systematic errors in the calculation of the nuclear matrix element. 

Finally, our result for the S-factor is,
\begin{equation}
 \label{eq_fineal_pp1}
 S^{11}(0)=(4.14\pm 0.01 \pm 0.005 \pm 0.06)\cdot 10^{-23}\text{ MeV fm}^2 .
 \end{equation} 
The quoted uncertainties are due to the different, in tension, measurements of $g_A$, the axial form factor of the nucleon~\cite{gA_1.2701, doi:10.1063/1.4983578} 
(see supplementary materials),
experimental uncertainties in the $^3$H decay, and the theoretical uncertainty. 
When used in solar simulations and studies, this number should be additionally corrected to include higher order electromagnetic effects, that are estimated at a 0.84\%~\cite{Acharya:2016kfl,Acharya:2019zil}, changing the central value to $S^{11}_{corr}(0)=4.11\text{ MeV fm}^2 $.
The current work continues the upward trend from the recent review of solar cross sections~\cite{RevModPhys.83.195}. The S-factor we find is consistent, albeit larger, with recent state-of-the-art chiral EFT calculations~\cite{Acharya:2019fij,Acharya:2016kfl,Acharya:2019zil}, as well as PhysRevLett.119.062002 QCD calculation of $L_{1,A}$~\cite{PhysRevLett.119.062002}. 

Our result for the S-factor represents an increase of $1-4\%$ over the former accepted proton-proton S-factor. The dominant role $pp$ fusion has in solar evolution entails that this creates a substantial effect on solar observables. For example, using the logarithmic sensitivities calculated in Ref.~\cite{Vinyoles:2016djt}, one finds that such an increase induces a reduction in the expected neutrino fluxes originating in $^8$B ($^7$Be) weak reactions in the Sun by $3-11\%$ ($1-4\%$) respectively. Substantial changes are expected in the future measurement of neutrinos from CNO cycle weak reactions. In particular, with regards to the puzzling tension between different determinations of solar photospheric elemental composition, i.e., the ``Solar Composition Problem'',~
\cite{2005ASPC..336...25A,doi:10.1146/annurev.astro.46.060407.145222,Magg2022,Asplund2021}, the increased value of the proton-proton fusion, and the consequent inferred neutrino fluxes, represent a better agreement with Ref.~\cite{Magg2022}, and a worse agreement to the analysis Ref.~\cite{Asplund2021}.

The extension of the current work to higher orders in \pilesseft is of importance to additionally confirm the uncertainty estimate. Previous studies of $pp$-fusion have reached N$^4$LO~\cite{Proton_Proton_Fifth_Order}, but their accuracy has been limited by the need to calibrate low-energy constants, e.g., $l_{1,A}$. Their result is indeed within the uncertainties of the current work. However, a theoretical approach using three body reactions to fix these LECs, thus increasing the calculation's accuracy, is still missing at orders higher than NLO. The challenge is mainly due to the need to include additional three-body forces and currents. This, as well as a systematic study of the effect of the calculated S-factor on Solar observables, is postponed to future work.

\begin{acknowledgments}
We thank S.\ K\"onig for sharing and benchmarking nuclear wave functions, and L.\ Platter, J.\ Vanasse, and J.\ Kirscher for
very helpful discussions. Research was
supported by the ISRAEL SCIENCE
FOUNDATION (grant No. 1446/16).
\end{acknowledgments}


\setcounter{equation}{0} 
\setcounter{table}{0} 
\setcounter{section}{0} 
\setcounter{subsection}{0} 
\setcounter{figure}{0}
\renewcommand{\thesection}{S}
\renewcommand{\theequation}{S-\arabic{equation}}
\renewcommand{\thesubsection}{S.\Roman{subsection}}
\renewcommand{\thefigure}{S.\arabic{figure}}
\renewcommand{\thetable}{S.\arabic{table}}

\section{supplementary Material}
\subsection{Impact of the $g_A$ uncertainty on $S^{11}(0)$}

As mention above, our source for $l_{1,A}$ calibration is the $^3$H $\beta$ decay. In \cite{de2019tritium} we calculated the $^3$H $\beta$ decay up yo NLO using the $\langle \|GT\rangle $ weak transition matrix element.
Up to NLO, the $\langle \|GT\|\rangle $ weak transition of $^3$H $\beta$ decay matrix element is written as:
\begin{eqnarray}
 &&\langle \|GT\|\rangle =\langle \|GT\|\rangle ^{\text{LO+NLO}}=\\
 \nonumber
 &&\langle \|GT\|\rangle ^{\text{LO}}+\langle \|GT\|\rangle ^{\text{NLO}}_{{l_{1,A}=0}}+l_{1,A}\times \langle \|GT\|\rangle ^{\text{NLO}}_{{l_{1,A}}}
\end{eqnarray}
where$\langle \|GT\|\rangle ^{\text{NLO}}_{{l_{1,A}}}$ are the two-body diagrams that contribute to the $^3$H $\beta$ decay and are coupled to l1,A, while
$\langle \|GT\|\rangle ^{\text{NLO}}_{{l_{1,A}=0}}$ is the sum over all the NLO diagrams that contribute to the $^3$H $\beta$ decay without the diagrams coupled
to $l_{1,A}$.
In Ref~\cite{de2019tritium} we found that the numerical value of $\langle\|GT\| \rangle$ is (for
 the Z-parametrization):
\begin{eqnarray}\label{eqn:GT_NLO}
 &&\langle \|GT\|\rangle =\langle \|GT\|\rangle ^{\text{LO+NLO}}=\\
 \nonumber
 &&\sqrt{3}\left[\underbrace{0.979}_{\text{LO}}-\underbrace{0.057}_{\text{NLO, }l_{1,A}=0}+l_{1,A}\times \underbrace{0.6}_{\text{NLO, }l_{1,A}}
\right]
\end{eqnarray}
Comparing Eq.~\ref{eqn:GT_NLO} to $\langle \|GT\|\rangle$ empirical number,
$
 \langle \|GT\|\rangle_{\text{emp}}=\sqrt{3}\frac{1.213\pm0.002}{g_A}$, one find that:
 \begin{equation}\label{eq:l_1A}
 l_{1,A}=\frac{1.213\pm0.002 - 0.921\times g_A}{0.6 \times g_A}
 \end{equation}
which is a function of $g_A$.

In the last decay, there were many estimations for $g_A$, and all of them are consistently higher than $g_A=1.2695$, the value indicated in Ref. \cite{RevModPhys.83.195,gA_2020}. Hence we can write $g_A$ as:
$g_A=1.2695(1+\epsilon)$, such that eq.~\ref{eq:l_1A} becomes (up to $\epsilon^2$):
\begin{equation}
 l_{1,A}=0.0575-1.59\epsilon+1.59\epsilon^2~,
\end{equation}
and, as a results, eq.~\ref{eq:Lambda_pp_calc} becomes:
\begin{equation}\label{eq:l1a_epsilon}
 \Lambda_{pp}(0)\approx 2.655+0.6\times l_{1,A}=2.6895-0.95\epsilon+0.95\epsilon^2
\end{equation}
However, for $S^{11}(0)$ (eq.~\ref{eq_S_11}, we have an extra term, $\frac{g_A^2}{1.2695.^2}$, which add an additional dependence on $g_A$. Hence, by combining eq.~\ref{eq_S_11} and \ref{eq:l1a_epsilon}, one can write $S^{11}(0)$ as a function of $\epsilon$ as: 
\begin{equation}
 S^{11}(0)=4.088\times(1+1.29\epsilon+0.42\epsilon^2)\mev\fm^2~,
\end{equation}
i.e., for any given calculation, the 0.48\% increase in $g_A$ from 1.2695 to 1.2756 \cite{gA_2020} results in a $0.62\%$ increase in $S^{11}(0)$. 
\begin{figure}[h]
 \includegraphics[width=1\linewidth]{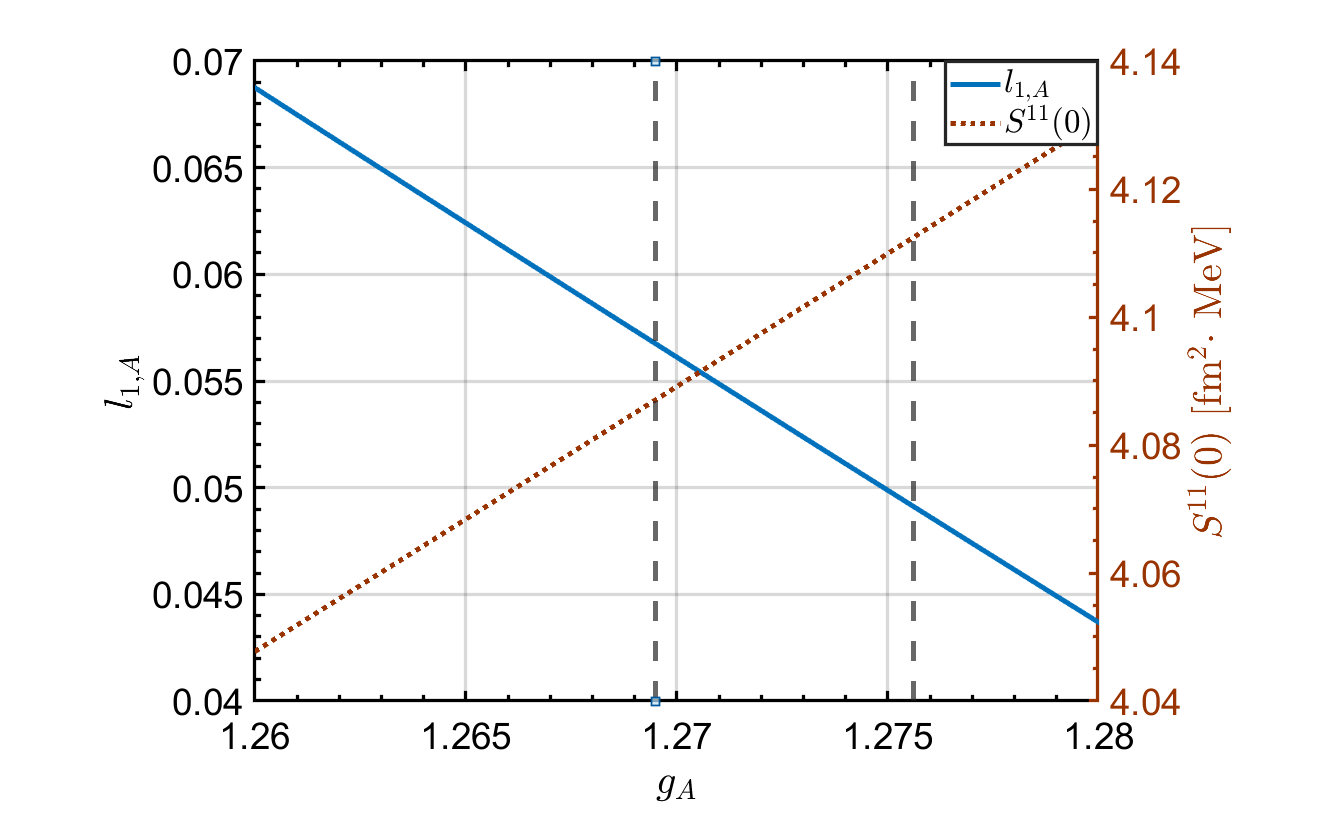}
 \caption{\textbf{The numerical calculation for $l_{1,A}$ and $S^{11}$}. Solid line: $l_{1,A}$; Dotted line: $S^{11}(0)$. The horizontal vertical dashed lines denote the values of $l_{1,A}$ and $S^{11}(0)$ for $g_A=1.2695$ (left) and $g_A=1.2756$ (right). \label{fig:l1a_pp}}
 \end{figure}brig
The numerical calculation for $l_{1,A}$ and $S^{11}$ are showing in Fig.~\ref{fig:l1a_pp}, where the solid line is the numerical results for $l_{1,A}$ and the dotted line is numerical results for $S^{11}(0)$. The horizontal-vertical dashed lines denote the values of $l_{1,A}$ and $S^{11}(0)$ for $g_A=1.2695$ (left) and $g_A=1.2756$ (right).

\bibliography{bibliography,reference}

\end{document}